\PassOptionsToPackage{svgnames}{xcolor}
\documentclass{vgtc}                          

\graphicspath{{figures/}{pictures/}{images/}{./}} 

\usepackage{times}                     
\usepackage{amssymb}
\usepackage{tabu}                      
\usepackage{booktabs}                  
\usepackage{lipsum}  
\usepackage{CJKutf8}
\usepackage{mwe}                       
\usepackage{amsmath}                    
\usepackage{mathptmx}                  
\usepackage{graphicx} 
\usepackage{epstopdf}
\usepackage{svg}
\usepackage{multirow}
\usepackage [english]{babel}
\usepackage [autostyle, english = american]{csquotes}
\usepackage[final]{changes} 

\definechangesauthor[name={adder}, color=blue]{+} 
\definechangesauthor[name={deleter}, color=red]{-} 
\newcommand{\ADD}[1]{\added[id={+}]{#1}}
\newcommand{\DEL}[1]{\deleted[id={-}]{#1}}

\MakeOuterQuote{"}

\onlineid{1032}

\vgtccategory{Research}

\vgtcinsertpkg

\title{Multimodal Feedback for Handheld Tool Guidance: Combining Wrist-Based Haptics with Augmented Reality}

\author{
  Yue Yang\thanks{first author, e-mail: yueyang1@stanford.edu}\\
  \scriptsize Stanford University
  \and
  Christoph Leuze\thanks{e-mail: cleuze@stanford.edu}\\
  \scriptsize Stanford University
  \and
  Brian Hargreaves\thanks{e-mail: bah@stanford.edu}\\
  \scriptsize Stanford University
  \and
  Bruce Daniel\thanks{e-mail: bdaniel@stanford.edu}\\
  \scriptsize Stanford University
  \and
  Fred Baik\thanks{corresponding author, e-mail: fbaik@stanford.edu}\\
  \scriptsize Stanford University
}

\abstract{
We investigate how vibrotactile wrist feedback can enhance spatial guidance for handheld tool movement in optical see-through augmented reality (AR). While AR overlays are widely used to support surgical tasks, visual occlusion, lighting conditions, and interface ambiguity can compromise precision and confidence. To address these challenges, we designed a multimodal system combining AR visuals with a custom wrist-worn haptic device delivering directional and state-based cues. A formative study with experienced surgeons and residents identified key tool maneuvers and preferences for reference mappings, guiding our cue design. In a cue identification experiment (N=21), participants accurately recognized five vibration patterns under visual load, with higher recognition for full-actuator states than spatial direction cues. In a guidance task (N=27), participants using both AR and haptics achieved significantly higher spatial precision (5.8 mm) and usability (SUS = 88.1) than those using either modality alone, albeit with modest increases in task time. Participants reported that haptic cues provided reassuring confirmation and reduced cognitive effort during alignment. Our results highlight the promise of integrating wrist-based haptics into AR systems for high-precision, visually complex tasks such as surgical guidance. We discuss design implications for multimodal interfaces supporting confident, efficient tool manipulation.
} 


\begin{document}


\firstsection{Introduction}
\maketitle
Surgeons routinely perform delicate procedures using handheld tools, such as scalpels, needles, and probes, where precise tool movement is critical for surgical success. Mastering such tool motions requires intensive training and experience to develop accurate hand-eye coordination. Indeed, surgical outcome quality has been shown to correlate with the surgeon’s experience: less experienced surgeons or those with low case volumes tend to have higher rates of suboptimal results in certain procedures \cite{birkmeyer2003surgeon, halm2002volume}. For example, in total knee replacement surgery, inadequate control of the surgical saw to locate cutting points on the bone can lead to improper implant alignment, and studies have found that less experienced surgeons are more likely to produce suboptimal alignments than experts \cite{kelz2021national, kraus2025fellowship}. This motivates the need for guidance systems that can assist surgeons in executing precise tool movements.

Recent technological advances have introduced tracking and navigation systems to guide tool positioning in surgery. Stereoscopic optical tracking cameras can localize instruments relative to patient anatomy in real time \cite{nema2022surgical, tsui2023optical}. However, these image-guided systems typically display information on an external 2D monitor, forcing the surgeon to look away from the operative field. This divided attention can disrupt focus, increase mental workload, and degrade hand-eye coordination \cite{ghazanfar2015effect}. 

Augmented reality (AR) has emerged as a promising solution for providing in situ guidance by overlaying virtual cues directly onto the surgeon’s view of the operative site \DEL{\cite{duan2025localization}}. For instance, AR guidance using inside-out infrared tracking reduced angular uncertainty during directional bone drilling, resulting in improved accuracy—benefits consistently seen across surgeons of varying experience levels \cite{van2024augmented}. In CT-guided interventions, a phantom study demonstrated that HoloLens 2–based AR reduced the number of needle pass redirections by approximately 54\%, cut radiation dose by roughly 41\%, and reduced procedure time by about 50\% \cite{park2020augmented}. In neurosurgical applications, AR-guided external ventricular drain (EVD) placement in a prospective clinical pilot achieved significantly higher first-attempt success rates, fewer revisions, and lower complication rates compared with conventional freehand techniques \cite{van2022high}.

However, exclusive reliance on visual AR cues also has limitations in dynamic and cluttered surgical environments. Depth perception remains problematic: misaligned overlays, insufficient occlusion, and perceptual conflicts all limit accuracy in near-field tasks \cite{condino2019perceptual,clinical}. Additionally, if the AR overlay occludes real anatomy or suffers registration errors, surgeons' trust can deteriorate \cite{doughty2022augmenting}. The already high visual burden of patient anatomy, instruments, and monitors may be exacerbated by AR, potentially increasing cognitive load or causing inattentional blindness \cite{doughty2022augmenting}. Thus, while AR provides valuable guidance, supplementing it with non-visual modalities could enhance robustness.

Haptic feedback offers an effective alternative channel, conveying guidance through touch and relieving visual demand. In clinical and experimental contexts, vibrotactile guidance serves as an intuitive “eyes-free” stimulus, helping maintain performance without overloading visual attention \cite{muzzammil2024review, aggravi2021haptic}. Wrist-mounted vibrotactile interfaces are particularly practical—they don’t obstruct the hands or line of sight and leverage the wrist’s sensitivity. Multi-actuator wristbands can deliver distinct directional cues (e.g., “rotate” or “advance”), enabling effective guidance \cite{rossa2016multiactuator}. Although spatial acuity remains limited with dense actuator arrays, these systems still support precise motion in many contexts. For example, Rossa et al. developed an eight-actuator wristband for needle steering in brachytherapy; surgeons recognized patterns with 86\% accuracy and executed correct adjustments 72\% of the time, even under high visual load \cite{rossa2016hand}.

Recent works have explored the integration of AR and haptic feedback. For instance, Yamamoto et al. (2012) developed an interoperable interface in robot-assisted laparoscopic surgery combining 3D visual overlays with haptic feedback, showing feasibility in tasks like palpation and surface tracing using virtual fixtures for safety \cite{yamamoto2012augmented}. Zhang et al. (2022) piloted a tool-based haptic guidance system for surgical navigation that reduced visual load and enabled users to perform path following and alignment tasks at sub-millimeter and sub-degree accuracy \cite{zhang2023towards}. These studies highlight the potential of multimodal guidance to enhance surgical precision while distributing information across sensory channels.

In this work, we explore the combination of AR and wearable haptic feedback to guide a handheld surgical tool. Our approach is motivated by the fact that many surgical manipulations occur relative to a reference plane – for instance, aligning an instrument along a bone surface or skin plane. Rather than guiding full 6-degree-of-freedom motions in free space (which can be cognitively demanding to follow with wrist-based haptics), we restrict guidance to planar directions on the reference surface (a 3-DOF in-plane translation, with separate cues for any required depth adjustment). We hypothesize that such tactile cues, when combined with AR visuals, can help the user navigate toward a target more efficiently than AR alone, especially under conditions of visual stress or occlusion.

To establish an intuitive “reference orientation” for these directional cues, we first conducted a formative study with experienced surgeons and residents (N=12). The goal was to determine what the “upward” direction of the tactile device should correspond to. We compared two mappings: (a) a fixed anatomical reference where the top (dorsal side) of the user’s wrist always corresponds to “up” on the device, versus (b) a tool-based reference where the device’s orientation is calibrated at the start (i.e., whichever way the tool is initially held defines the forward/up direction). The formative study (described below) indicated that a fixed wrist-referenced mapping was preferred for consistency. Using this mapping, we integrated the haptic device with an AR system (HoloLens 2) to track the tool and display visual targets. We then conducted two main experiments with a total of 48 participants. Experiment 1 (N=21) evaluated how well users can perceive and interpret the vibrotactile cues alone (direction and distance cues) in a lab setting. Experiment 2 (N=27) assessed the effectiveness of the tactile guidance in a simulated surgical targeting task on a knee phantom, comparing three conditions: haptic guidance only, AR visual guidance only, and combined AR+haptic guidance.

\section{Related Work}
\subsection{Haptic Guidance in Surgery and Training}
Previous studies indicate that vibrotactile haptic feedback can enhance surgical training by conveying guidance through touch. Wearable tactile devices, like vibration wristbands, have been used successfully in tasks such as guiding needle steering in brachytherapy \cite{raitor2023design, li2022review}. Moreover, meta-analytic evidence shows that haptic feedback in robot-assisted surgery significantly reduces average and peak applied forces, shortens task completion time, and increases accuracy and success rates \cite{bergholz2023benefits}. Additional research in surgical simulation demonstrates that haptic-enhanced training improves technical precision—such as reduced drill penetration and better structured assessment scores—compared to training without haptics. 

Our work builds on prior research into wearable haptic interfaces for directional guidance, particularly wrist-worn vibrotactile bands, prized for their convenient location, tactile sensitivity, and social acceptability \cite{hong2016evaluating, elsayed2020vibromap}. Early systems using a single actuator, such as a buzz on the left versus right wrist to indicate direction, achieved moderate guidance success ($\sim$73\% correct in one study) \cite{wang2018empirical}. To convey more nuanced directions, researchers have explored multi-actuator designs, arranging 4, 6, or 8 vibrators around the wrist or forearm to generate localized or moving sensations indicating travel direction or object location \cite{salazar2018path, rossa2016multiactuator}. Studies also show that dynamic patterns, such as sequential or alternating activation, improve recognition over static single-actuator signals \cite{hessinger2020wearable, elsayed2020vibromap}. We build on these findings by using 12 vibrators plus phantom interpolation, achieving about 24 virtual directions ($\sim$15° resolution).

\subsection{Augmented Reality and Multimodal Guidance}
Emerging multimodal techniques pair AR with haptic feedback—especially wrist-worn or tool-integrated vibrotactile and kinesthetic cues—to maintain directional guidance even when visuals degrade. A compelling example is a pilot study where a surgical tool’s handle included vibrotactile actuators \cite{zhang2023towards}. Participants, guided by haptic feedback alone, successfully traced hidden paths and aligned instruments with sub-millimeter and sub-degree accuracy. This suggests robust spatial guidance can be achieved without visual support. 

In a different context, HapticAR introduces a wrist-worn vibrotactile system that complements AR overlays during assembly tasks \cite{arbelaez2019haptic}. This setup reduced reliance on visual instructions, improved part placement accuracy, and alleviated cognitive load. The Tasbi wristband expands upon this idea by offering both squeeze and vibrotactile feedback, delivered via six radially distributed actuators \cite{pezent2019tasbi}. Engineered for AR/VR environments, Tasbi enhances tactile responsiveness while keeping users’ hands free for interaction. For more visceral interaction scenarios, a wearable, tendon-driven device provides kinesthetic stiffness rendering during AR finger manipulation \cite{lee2021wearable}. This design prevents interpenetration with virtual objects and offers realistic tactile sensation within minimal occlusion and compact form.

Broader reviews of haptic AR underscore its potential to enrich real object perception—especially when visual rendering alone is insufficient—highlighting haptic augmentation as critical for immersive and accurate spatial interaction \cite{bhatia2024augmenting}. Experiments integrating visual arrows and haptic prompts in dual-task scenarios revealed that multimodal guidance improves reaction time and user experience, allowing users to maintain focus on a primary task while effectively responding to haptic cues \cite{Wang_Wang_Ren_2024}. Our work is, to our knowledge, one of the first to evaluate an AR + wrist haptic guidance system in a surgical targeting scenario for tool guidance. By integrating continuous tactile navigation cues with AR visual cues, we aim to combine the strengths of each: the intuitive spatial context of AR with the eyes-free directional prompting of haptics.

\section{Formative Study: Determining Reference Orientation and Defining Tool Maneuvers}

Before finalizing the haptic guidance design, we conducted a formative study to determine two aspects of the vibrotactile feedback system: (1) the most intuitive reference orientation for directional cues, and (2) a set of common tool maneuvers that surgeons routinely perform, which could be encoded into distinct haptic patterns.

\begin{figure}[h!]
\begin{centering}
    \includegraphics[width=\linewidth]{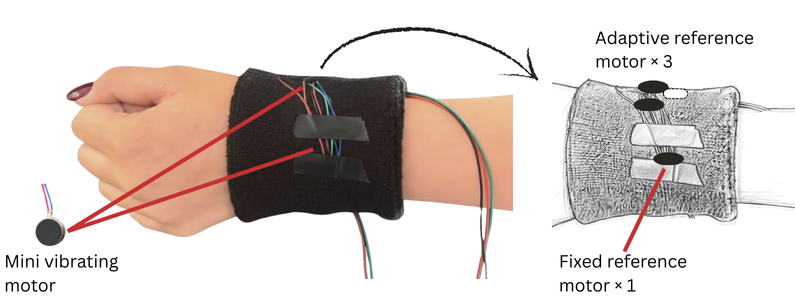}
    \caption{\ADD{The custom haptic wristband used in the formative study embedded vibrotactile motors for direct skin contact. Its flexible design conformed to different wrist sizes. Three motors defined an “up” direction relative to gravity, with one selected based on initial tool orientation, while a fourth motor defined a wrist-relative “up” direction that remained constant across participants.}}
    \label{fig:1}
    \end{centering}
\end{figure}

We developed a custom vibrotactile wristband controlled via an Arduino Mega microcontroller. The device comprises four mini eccentric rotating mass (ERM) vibratory actuators (10~mm diameter, 3 mm thickness), each independently controllable \cite{Industries}. As shown in \cref{fig:1}, three actuators are positioned in a “tool-oriented, adaptive-reference” location, while one is fixed in a “wrist-up” reference location. 
For the adaptive placements, the actuator oriented upward relative to gravity—determined by the user's initial tool grasp—serves as the reference “up” position. This configuration enables evaluation of the two candidate mappings described in \cref{sec:3.1}.

We selected ERM disc-type motors due to their ubiquity in haptics research, low cost, and ease of procurement. Their flat form factor allows straightforward integration into a wristband design. Similar motors (10~mm circular ERMs) have been previously employed effectively in wrist-worn prototypes for eliciting phantom tactile sensations and spatial feedback \cite{hong2016evaluating, scheggi2014vibrotactile}. To enhance tactile sensation and minimize inter-actuator vibration transmission, each motor was embedded directly into the wristband to ensure direct skin contact, following prior design considerations \cite{rossa2016multiactuator}. 

Each motor is connected to and managed through its own relay interface, enabling independent activation. All motors vibrate at a fixed frequency of 150 Hz, close to their mechanical resonance, optimizing skin stimulation efficacy. This frequency choice aligns with findings that highlight the effective and perceptually salient vibration range for human skin, particularly within the Pacinian corpuscle sensitivity spectrum (typically 150–300 Hz) \cite{martinez2022psychophysical}.

\subsection{Reference Orientation for Vibrotactile Cues} \label{sec:3.1}
\begin{figure}[t]
\begin{centering}
    \includegraphics[width=\linewidth]{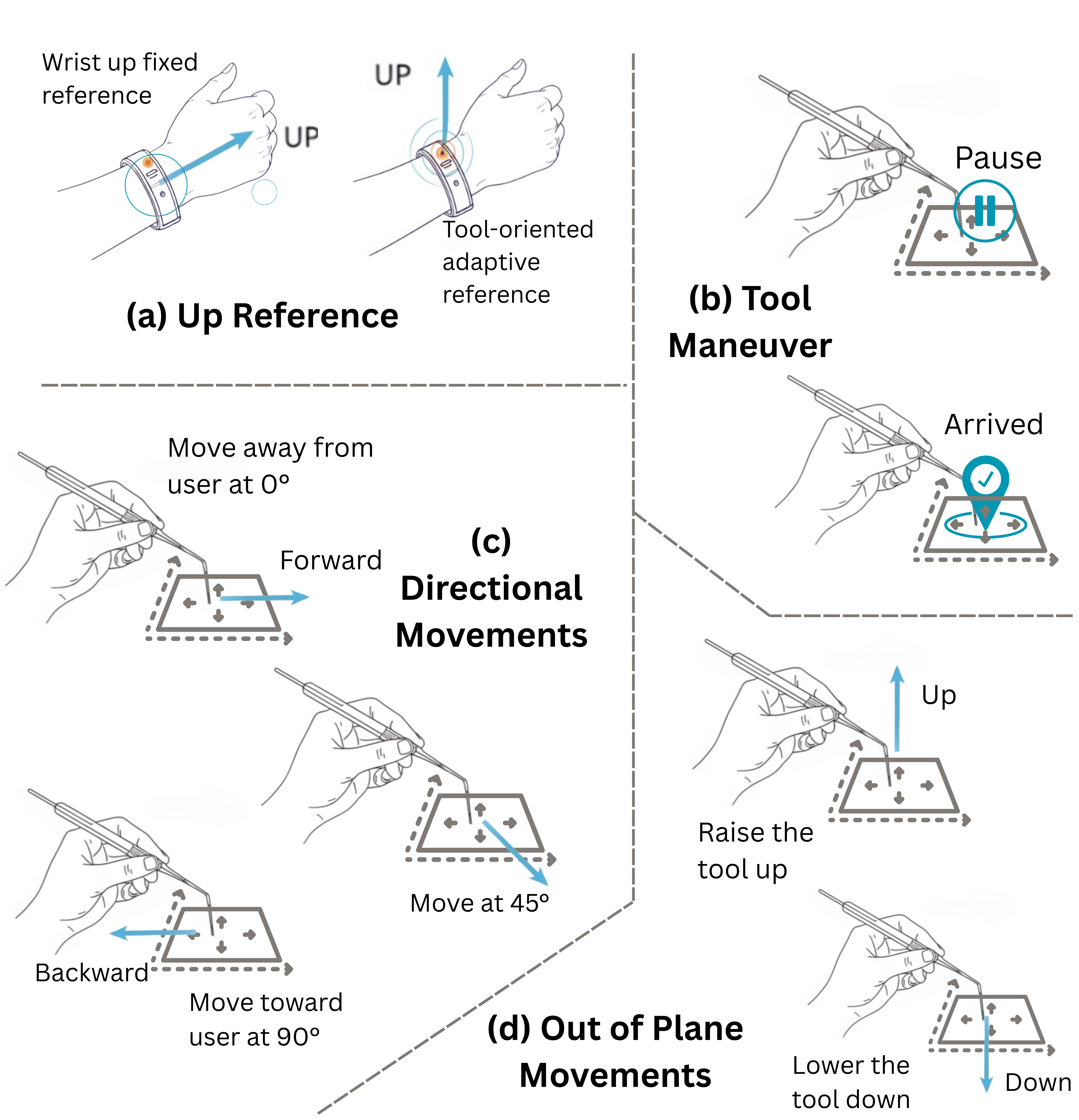}
    \caption{\ADD{Reference orientations and common tool movements from the formative study. (a) compares fixed versus adaptive reference frame mappings. (b–d) depict tool maneuvers, directional movements, and out-of-plane motions identified by users, which informed the vibration pattern design and subsequent experiments.} }
    \label{fig:2}
    \end{centering}
\end{figure}

A central question in the design of directional cues was how to define the “upward” direction of the vibration signals, given that users may hold the tool at arbitrary orientations. As shown in \cref{fig:2}a, we considered two candidate mappings:

\begin{itemize}
    \item Wrist-Up (Fixed Reference Frame): The “up” direction corresponds to the dorsal (top) side of the user’s wrist, regardless of tool orientation. Thus, a vibration at the top of the wrist always signifies an upward cue within the reference plane, anchoring the mapping to the user’s anatomy.
    \item Tool-Oriented (Adaptive Reference Frame): The orientation is calibrated based on the user’s initial grip. At the beginning of the session, whichever actuator faces upward relative to gravity is defined as “up.” Subsequent cues are tied to this frame, which may shift if the tool is rotated.
\end{itemize}

\ADD{After informed consent, 12 participants (26–55 years,  nine male and three female, four experienced surgeons and eight surgical residents) evaluated both mappings and finished a semi-structured post-interview.} Each participant wore the haptic wristband on their dominant hand while holding a mock surgical tool \cite{Industries}. For each of the two mapping conditions, participants were presented with a set of 20 discrete vibrations, each lasting 0.5 seconds. Upon receiving each cue, participants were instructed to interpret the signal and move the tool tip straight away from them in a forward direction (see \cref{fig:2}c) along the table plane. They were prompted to rotate the tool freely to explore how cues aligned under each reference frame.

Nine of the 12 participants expressed their preference for the wrist-up mapping. They noted that this anatomy-referenced coordinate system offered a stable and predictable experience—for example, a vibration on the dorsal side of the wrist consistently signaled a “move forward” command, regardless of how the tool was held. In contrast, the tool-oriented mapping was frequently described as confusing, especially when the tool was rotated mid-task, causing the directional cues to shift in ways that conflicted with users’ mental models. Although participants gradually became more comfortable with the tool-oriented scheme over time, many noted that it required a steeper learning curve. As one participant mentioned, “At first it felt like the vibrations were all over the place, but after some tries, it started to click.” Another participant also mentioned, “It [tool-oriented] eventually made sense, but wrist-up just worked from the start.” The remaining three participants, who were neutral, adapted to it more quickly, but still acknowledged the intuitiveness of the wrist-up approach. Given the overall consensus, the wrist-up mapping was chosen for all subsequent experiments and formed the foundation for our vibration sequence and cue design.

\begin{figure*}[tbh]
\begin{centering}
    \includegraphics[width=\linewidth]{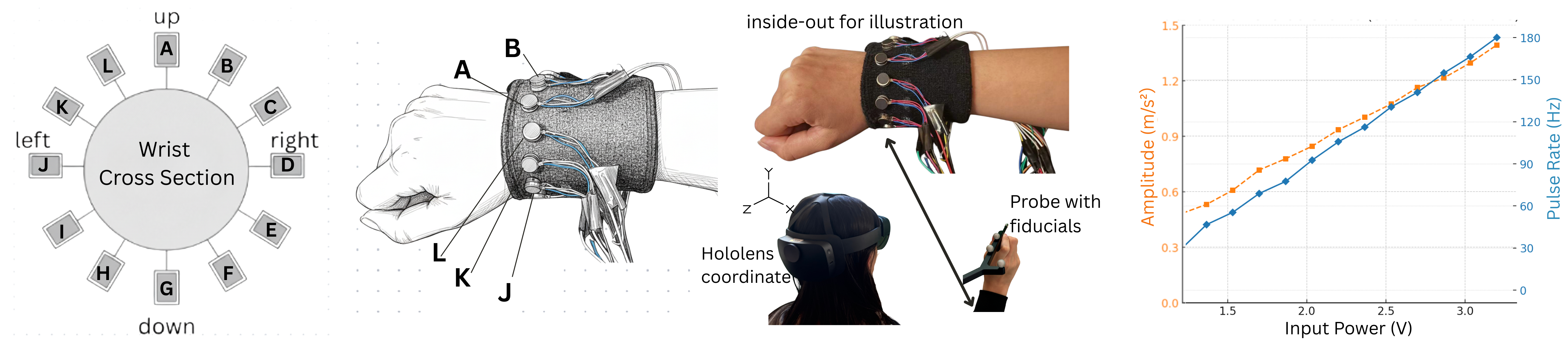}
    \caption{Haptic system design. A wristband with 12 individually controlled vibrotactile motors was driven by an Arduino microcontroller, which communicated with the HoloLens via serial connection. Tool motion was tracked using the HoloLens depth camera, and vibration amplitude and frequency were measured with an embedded accelerometer. \ADD{Amplitude is in 9.81 m/s² (1 gravitational acceleration).}}
    \label{fig:3}
    \end{centering}
\end{figure*}

\subsection{Definition of Common Tool Movements and Maneuvers}
In addition to evaluating reference orientation, participants were asked to define the set of tool movements and maneuvers they considered most common in surgical practice during the interview. As illustrated in \cref{fig:2}b to d, the following categories were identified through discussion and consensus: 1) \textbf{Discrete tool maneuvers.} Two specific tool actions frequently used in surgical contexts: \textit{Pause} (temporarily holding the tool position without further movement) and \textit{Arrived} (reaching the intended target location); 2) \textbf{Planar directional movements.} Tool movements along any direction along the operating plane. Three example directions are shown in \cref{fig:2}c; 3) \textbf{Out-of-plane movements.} Two motions perpendicular to the reference plane (\textit{Raise \& Move Up} or \textit{Lower \& Move down}).

These definitions informed the design of the vibrotactile cue sequence, enabling the system to encode both continuous directional guidance (planar or out-of-plane) and discrete maneuver commands.

\section{Haptic System Design} \label{sec:4}

Driven by insights from our formative study, we developed a custom haptic feedback device to guide the identified tool directional movements and discrete maneuvers. As shown in \cref{fig:3}, the device comprises 12 miniature eccentric rotating mass (ERM) vibration motors (10~mm diameter) evenly spaced in 30° intervals around a soft wristband to ensure full angular coverage. The decision to use 12 motors was informed by prior work demonstrating that increasing motor count from 4 to 8 reduced directional movement errors in fingertip-based tasks \cite{hong2016evaluating}, and that 12 motors provide near-continuous coverage around the wrist.

Each motor is controlled independently using pulse-width modulation (PWM) via an Arduino Mega microcontroller, allowing for precise control of vibration amplitude. To synchronize the motor actuation with tool movement, we developed a custom Unity (v2022.3.8) application on the HoloLens 2. This application communicates with the Arduino board via the serial communication protocol. The handheld tool is equipped with IR fiducials \ADD{(infrared markers visible to the tracking camera)} and tracked in real time using the HoloLens 2’s onboard depth camera through a marker tracking algorithm adapted from \cite{martin2023sttar}. The same IR fiducials and tracking algorithm were used for target registration on the operating plane, synchronizing everything into the HoloLens coordinate system. All motors are embedded directly within the wristband to maintain close skin contact for effective haptic perception.

\subsection{Phantom Sensation} \label{sec:4.1}
Our haptic design addresses a critical challenge that tool movements can occur in any direction within 360°, necessitating high angular resolution. To address this, we leverage phantom sensation (a perceptual illusion where simultaneous vibration of two adjacent actuators is perceived as a single vibration) \cite{kato2010basic}. Prior studies suggest that amplitude balance between motors strongly influences the quality of the phantom sensation \cite{remache2024phantom, alles1970information}. To define appropriate vibration thresholds, we empirically characterized each motor's amplitude and frequency response across a voltage sweep using an embedded accelerometer. As shown in \cref{fig:3}, we set the lower and upper amplitude bounds to $0.3\,\mathrm{m/s^2}$ and $1.3\,\mathrm{m/s^2}$ (or $1.3\,\mathrm{V}$ and $3\,\mathrm{V}$), respectively.

To generate phantom sensations for non-exact target angles, we applied a linear interpolation strategy to compute the voltage levels delivered to the two neighboring actuators. This choice was informed by previous findings showing that linear interpolation produces more perceptually accurate phantom sensations than logarithmic schemes \cite{rahal2009investigating, schafer2025vibrotactile, kirchner2023phantom}. The voltage values are determined using the following equations:

\[
\begin{aligned}
V_{\alpha} &= \left( \frac{\phi_{R} - \phi - 10}{\phi_{R} - \phi_{L} - 10} \right)(V_{\text{max}} - V_{\text{min}}) + V_{\text{min}} \\
V_{\beta} &= \left( \frac{\phi - \phi_{L} - 10}{\phi_{R} - \phi_{L} - 10} \right)(V_{\text{max}} - V_{\text{min}}) + V_{\text{min}}
\end{aligned}
\]

In these expressions, \( V_{\alpha} \) and \( V_{\beta} \) represent the voltages applied to the two motors neighboring the target direction \( \phi \), with \( \phi_{L} \) and \( \phi_{R} \) indicating the angular positions (in degrees) of the left and right neighboring motors, respectively, such that \( \phi_{L} < \phi < \phi_{R} \). The parameters \( V_{\text{min}} = 1.3\,\text{V} \) and \( V_{\text{max}} = 3\,\text{V} \) denote the empirically determined minimum and maximum voltage bounds. The constant 10 reflects the angle between the 36-direction layout for our tasks in experiment 1. Interpolation is only performed when the angular distance between the target direction and its nearest motor exceeds 10°; otherwise, a single motor is activated at maximum voltage. \ADD{When two neighboring motors were driven simultaneously, users consistently reported a clear single "virtual" vibration. This provided smoother feedback, although the 12 motors by themselves  decent resolution.}

\begin{figure*}[tbh]
\begin{centering}
    \includegraphics[width=\linewidth]{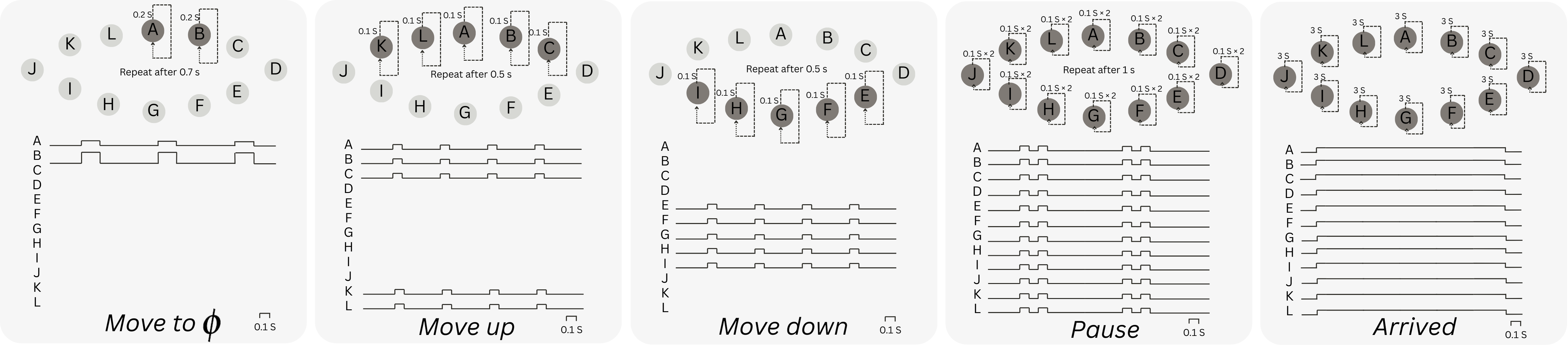}
    \caption{Vibrotactile patterns encoding five tool actions. We designed five haptic cue sequences to guide tool movement during surgical tasks: (a) planar directional movement, or \textit{Move to}, (b) \textit{Move up}, (c) \textit{Move down}, (d) \textit{Pause}, and (e) \textit{Arrived}. Continuous movement cues—planar and out-of-plane—varied in amplitude, frequency, actuator count, and inter-pulse intervals. Out-of-plane cues were triggered when tool position deviated more than ±5 mm from the defined movement plane, activating five top-side (up) or bottom-side (down) actuators. Upon re-entering the valid plane, a discrete pause signal was issued using two bursts from all actuators. When the tool reached within 10~mm of the target center, a 3-second full-actuator signal confirmed successful arrival. Patterns were designed to be easily distinguishable and grounded in formative studies on surgical tool maneuvers.}
    \label{fig:4}
    \end{centering}
\end{figure*}

\subsection{Vibration Patterns}
We designed a set of vibrotactile patterns to serve as haptic cues, each signaling one of five predefined tool actions: planar directional movement, \textit{Move up}, \textit{Move down}, \textit{Pause}, and \textit{Arrived}, as illustrated in \cref{fig:4}. These actions were further grouped into two categories: (1) continuous motions, which include both planar and out-of-plane directional movements, and (2) discrete static maneuvers, including pause and arrival signals. Each haptic sequence was modulated by varying the vibration amplitude, duration, number of activated actuators, and inter-pulse intervals. The design of these cues was informed by findings from formative studies, which identified common tool maneuvers used in the operating room.

For planar directional movement, the “forward” direction (\cref{fig:2}c), defined as the wrist-up orientation (0°, \cref{fig:2}a), was used as the reference. To encode proximity to the target, pulse frequency increased as the tooltip approached the target center point, following a motion metaphor in which urgency grows with proximity \cite{erp2005waypoint}. Specifically, three distance categories were defined based on the tool tip's starting position: (1) \ADD{Far (\textgreater50 mm)}: 1.0 s pulse interval; (2) Medium (20–50 mm): 0.7 s; and (3) \ADD{Close (\textless20 mm)}: 0.4 s. Thus, users received slow, infrequent pulses when far from the target, with increasing frequency as they moved closer.

An example sequence is shown in \cref{fig:4}a, representing a planar directional movement toward a target located 20° from the forward direction and within the medium distance category. Because surgical tool movement often occurs within a defined 2D plane, we defined a tolerance zone of ±5 mm (10mm thick zone) in the out-of-plane direction as the acceptable “in-plane” margin. This error margin accommodates natural human tissue (e.g., skin or bone) bending, tool tracking inaccuracies, and spatial mapping errors (e.g., HoloLens SLAM drift) and is consistent with the precision shown in prior work on haptics-guided movement \cite{rossa2016multiactuator}.

Out-of-plane deviations were detected when the tool tip moved beyond this 10~mm zone. \ADD{As shown in \cref{fig:2}d}, deviations triggered vertical correction cues. For move-up signals, the top-side actuators (K, L, A, B, C; \cref{fig:4}b) vibrated. For move-down signals, the bottom-side actuators (I, H, G, F, E; \cref{fig:4}c) were activated. These vertical cues used a distinct pattern: five actuators buzzing for 0.1 s, repeated every 0.5 s, differentiating them clearly from planar directional signals. Once the tool returned to within the in-plane margin, a pause signal was triggered, followed by a resumption of planar guidance.

Pause and arrival cues were defined as discrete, one-time signals. The pause cue was triggered when the tool re-entered the valid plane after a vertical deviation. This was encoded as two rapid bursts from all 12 actuators (0.1 s on ×2), repeated once per second (\cref{fig:4}d), signifying a clear “stop” warning. The arrival cue, triggered when the tool tip was within 10~mm of the target center, consisted of a continuous 3-second vibration of all actuators (\cref{fig:4}e), conveying successful target acquisition. This distinctive “all motors” signal was deliberately designed to be unmistakable and celebratory, providing clear feedback that the task was completed.

\section{Experiment 1: Cue Identification Study}
\subsubsection*{Design} The first experiment evaluated participants’ ability to recognize movements and maneuvers corresponding to each haptic cue sequence, under the potential interference of tool tracking AR visual overlay. We systematically varied the vibration sequences and measured recognition accuracy, confusion rates, and response times. Response time was defined as the interval between the completion of the cue sequence and the participant’s verbal report. The presentation order of sequences was randomized in this within-subjects design. We additionally conducted statistical analyses to test for differences among patterns.  

We formulated two hypotheses: \textbf{H1:} Recognition rates would not differ significantly among the five haptic patterns. We hypothesized that the uniquely designed vibration patterns, distributed across 12 actuators, would result in no systematic advantage of one pattern over another. We also assumed that the AR overlay would not impact vibration sensation, causing minimal disturbance. \textbf{H2:} The \textit{Pause} and \textit{Arrived} cues would yield shorter response times compared with planar and out-of-plane directional movements, since both involve simultaneous activation of all motors, making them more salient and easier to identify.  

\subsubsection*{Participants} \ADD{21 right-handed participants (19-34 years, 15 male and six female)} were recruited from the university. None had prior experience with wrist-based haptic devices. All were students or staff members, and approximately 20\% reported prior participation in AR-related studies.  

\subsubsection*{Procedure} After providing informed consent and completing a demographic survey, participants received instructions on holding the tool, wearing the wristband and the HoloLens 2 HMD. As illustrated in \cref{fig:3}c, they were asked to hold the tool in a posture consistent with common surgical grips (e.g., scalpel handling). The haptic system described in \cref{sec:4} was used throughout the study.  

Participants sat at a desk with their right elbow resting on the surface. With the tool held naturally, the wristband was positioned near chest height. They were instructed to extend the tool at arm’s length and freely rotate the wrist during the experiment. The virtual proxy of the tool was continuously overlaid onto the physical tool by a HoloLens tracking algorithm \cite{martin2023sttar}, and this tool tracking visualization is displayed to participants across experiment 1. 

Two practice sessions were conducted. In the first, each pattern was played four times in random order, and participants verbally confirmed the corresponding action. For planar directional patterns, cues corresponding to up, down, left, and right were presented. In the second session, diagonal planar cues (45°, 135°, 225°, 315°) were introduced using the same procedure.  

The main session followed practice sessions, with each participant completing: 1) 24 trials of planar directional movements (in 15° increments),  and 2) 10 trials each of \textit{Move up}, \textit{Move down}, \textit{Pause}, and \textit{Arrived} cues. This resulted in 64 randomized trials per participant, covering all five cue categories. After each trial, participants verbally reported the recognized category (e.g., they may say ``up'' for \textit{Move up}, ``down'' for \textit{Move down}). They were not required to identify the exact angular direction for directional cues, as this was not the focus of Experiment~1; only categorical recognition was measured.

\subsubsection*{Results} In total, 1,344 trials (64 $\times$ 21 participants) were collected. The primary outcome was \textit{recognition rate}, defined as the proportion of correctly identified cues. Confusion rates were also calculated. In addition, we recorded 1,344 response times using manual logging: the study coordinator pressed the spacebar when a verbal response was heard. A custom Unity package with serial communication was used to control motor activation and timelog both trigger events and button presses.

\begin{figure}[tbh]
\begin{centering}
    \includegraphics[width=\linewidth]{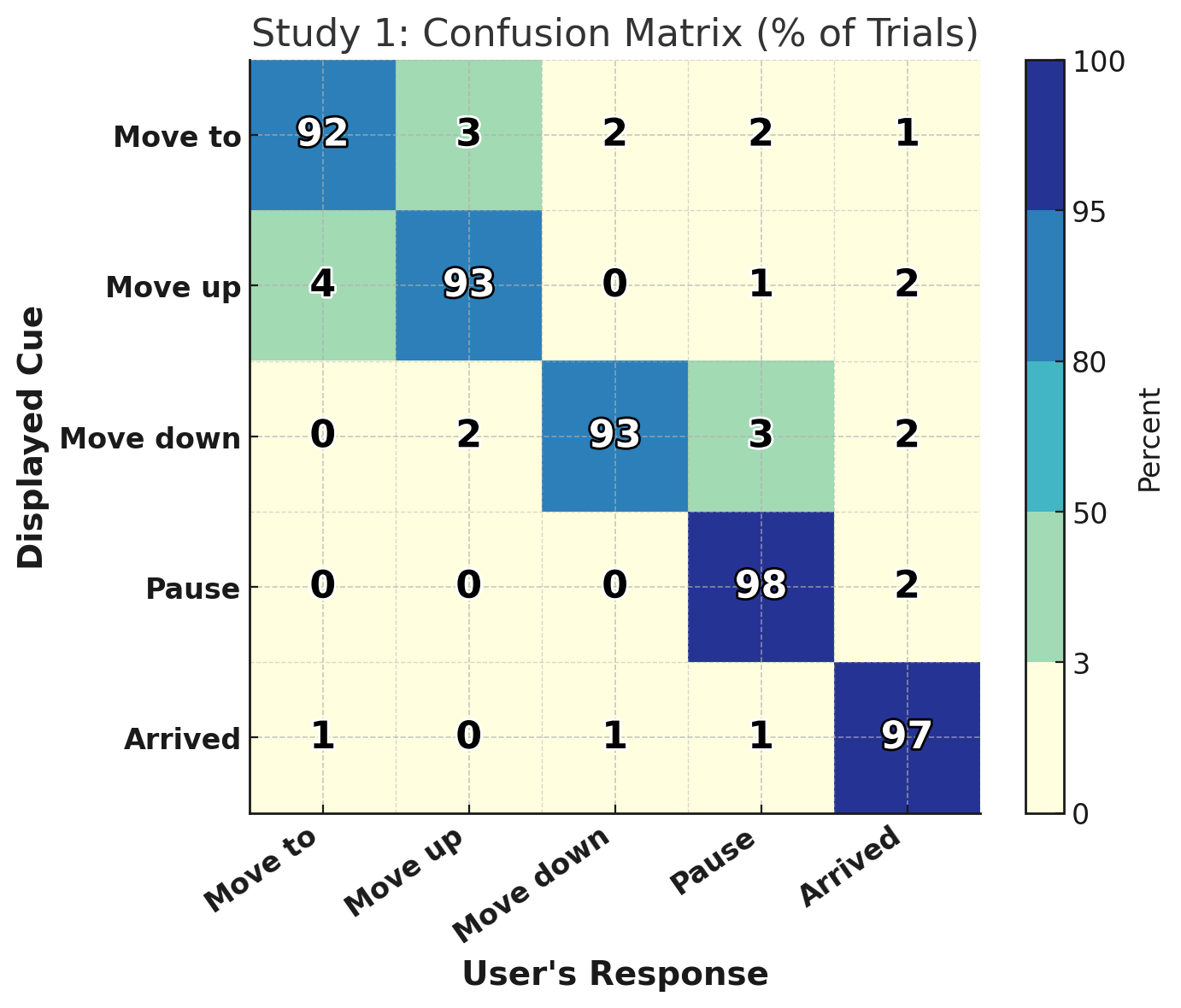}
    \caption{Mean recognition rates for the five haptic cue categories (\textit{Move to}, \textit{Move up}, \textit{Move down}, \textit{Pause}, and \textit{Arrived}), averaged across all participants. Darker colors indicate higher percentiles.}
    \label{fig:5}
    \end{centering}
\end{figure}

We first examined whether recognition rates differed among the five vibration patterns (\textit{Move to}, \textit{Move up}, \textit{Move down}, \textit{Pause}, \textit{Arrived}). Shapiro--Wilk tests indicated that recognition rates deviated from normality for all categories ($p < .05$). \ADD{In this dataset, the generally high performance across participants meant that between-subject variance was small.} Thus, we employed the non-parametric Friedman test. The Friedman test revealed a significant main effect of pattern category on recognition accuracy, $\chi^{2}(4) = 11.26, p = .024$, indicating that recognition performance varied across cue types. Descriptive statistics showed that recognition rates were high across all conditions, with mean accuracies of 92\% for \textit{Move to}, 93\% for \textit{Move up}, 93\% for \textit{Move down}, 98\% for \textit{Pause}, and 97\% for \textit{Arrived}.

Post-hoc Nemenyi test revealed that both \textit{Pause} ($M = 98\%$) and \textit{Arrived} ($M = 97\%$) were recognized with significantly higher accuracy than the three directional cues (\textit{Move to}, \textit{Move up}, and \textit{Move down}; all adjusted $p < .05$). No significant differences were observed among the three directional categories ($p > .10$).

These results indicate that cues activating all motors simultaneously (\textit{Pause} and \textit{Arrived}) produced more salient and distinctive tactile signatures, resulting in higher recognition accuracy. In contrast, directional cues required discrimination of localized stimulation patterns. This finding is noteworthy because our original hypothesis (H1) predicted no differences in recognition rate across categories. Instead, the results demonstrate that multi-motor activation leads to consistently more robust cue identification, indicating that the AR overlay may cause certain visual disturbances, which adversely impact the haptic recognition of less obvious vibrations from fewer motors.

Hypothesis~2 predicted that the \textit{Pause} and \textit{Arrived} cues would yield lower response times compared to directional cues, as these patterns activated all motors simultaneously and were expected to be easier to identify. Shapiro--Wilk tests revealed that response times were non-normally distributed for conditions ($p < .05$); therefore, we conducted a Friedman test to compare median response times across the five categories.

The Friedman test indicated no statistically significant main effect of pattern category on response time, $\chi^{2}(4) = 5.83, p = .21$. Nevertheless, descriptive statistics revealed consistent trends in line with our hypothesis: \textit{Pause} ($M = 2.18\,\text{s}, SD = 0.41$) and \textit{Arrived} ($M = 2.24\,\text{s}, SD = 0.39$) had the lowest average response times. By contrast, the three directional cues required longer response times: \textit{Move down} ($M = 2.71\,\text{s}, SD = 0.83$), \textit{Move up} ($M = 2.84\,\text{s}, SD = 0.49$), and \textit{Move to} ($M = 2.89\,\text{s}, SD = 1.52$). 

Although these differences did not reach statistical significance, the observed pattern supports H2: cues with full-motor activation (\textit{Pause} and \textit{Arrived}) were recognized more quickly, whereas directional cues imposed longer latencies. This suggests that, under AR visual disturbance, holistic stimulation patterns reduce cognitive processing demands and facilitate faster recognition, whereas localized directional cues are relatively more effortful to identify. Such a trend underscores the potential efficiency benefit of multi-motor activation in time-critical contexts.

\section{Experiment 2: AR and Haptics Guided Targeting Task}
\subsubsection*{Design}
In Experiment 2, we further evaluated the impact of wrist haptic feedback on actual task performance in a simulated tool guidance scenario for screw placement during total knee replacement surgery, where the knee bone is being exposed. The study employed a within-subjects design where each participant performed the task under three feedback conditions: Haptics-only, AR-only, and Haptics+AR combined. We then measured objective performance data (end-point deviation, time to completion) and subjective workload/usability in each condition. \ADD{We formulated two more hypotheses: \textbf{H3:} We hypothesize that combining AR visuals with haptic cues (Haptics+AR) will improve guidance performance with the highest precision and lowest time-to-target. \textbf{H4:} Haptics alone would show the highest workload with the lowest usability score.}

\begin{figure}[tbh]
\begin{centering}
    \includegraphics[width=\linewidth]{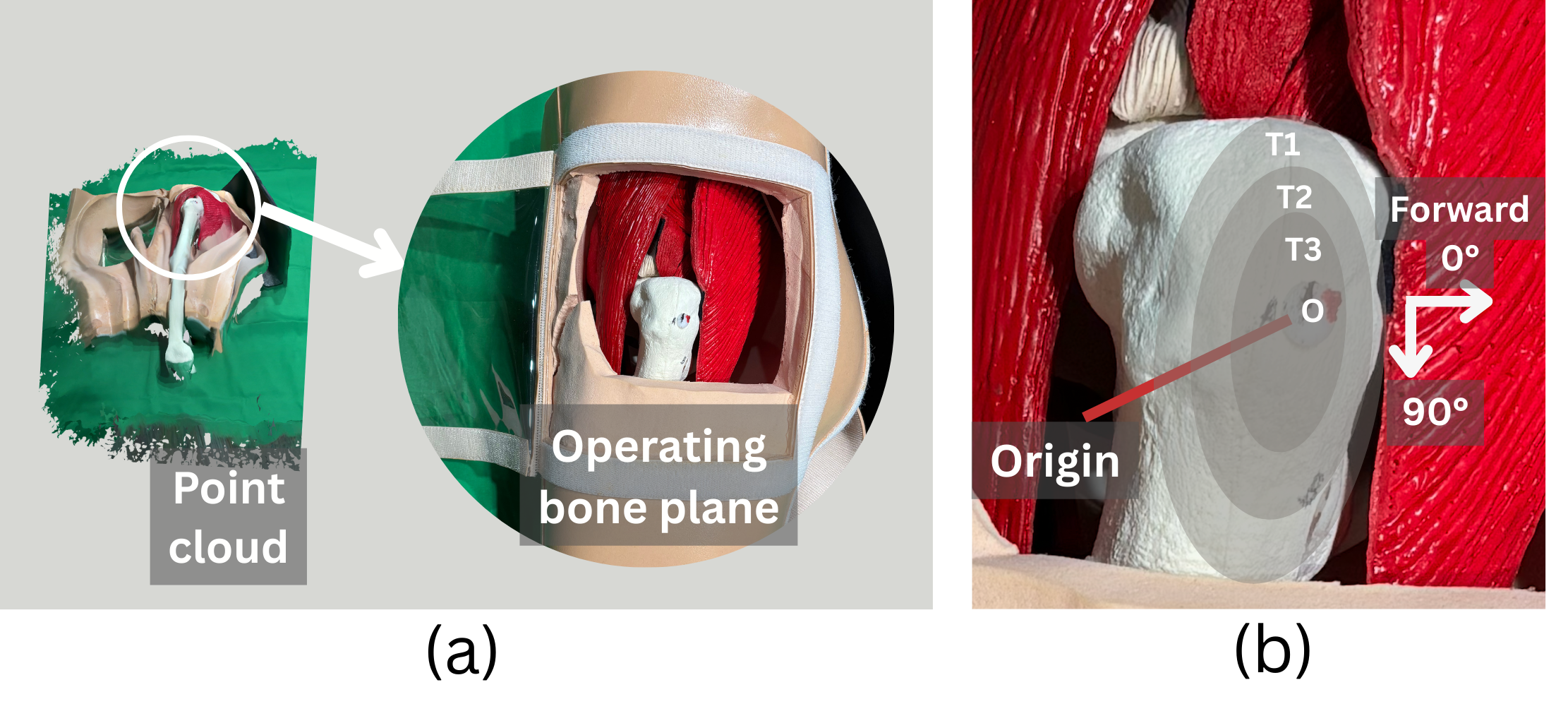}
    \caption{Experiment 2 study setup. (a) The knee phantom was fixed on a surgical desk, and a point cloud reconstruction was performed using the HoloLens depth camera as part of the markerless AR registration process. As shown in the zoomed-in image of the operating bone plane, to simulate surgical conditions, only the knee bone was visible during the experiment, reflecting realistic exposure of the target anatomy. (b) Illustration of the defined coordinate system, showing the forward (0$^{\circ}$) and lateral (90$^{\circ}$) directions, as well as the three radial target zones (T1, T2, T3) centered around the origin. The origin served as the standardized starting location of the tool tip at the beginning of each trial.}
    \label{fig:phantom}
    \end{centering}
\end{figure}

\subsubsection*{Participants}
\ADD{27 volunteers (19-53 years, 12 male and 15 female) were recruited from Stanford University and Stanford Medicine and provided informed consent,} \ADD{including eight surgeons and residents.} Six of them have participated in Study 1, who are tagged to analyze the impact of prior system training and familiarity on task performance. They were screened to ensure no significant impairment in tactile sensation (since the task relies on feeling vibrations). Around 30\% of them had prior experience with AR.

\begin{figure*}[tbh]
\begin{centering}
    \includegraphics[width=\linewidth]{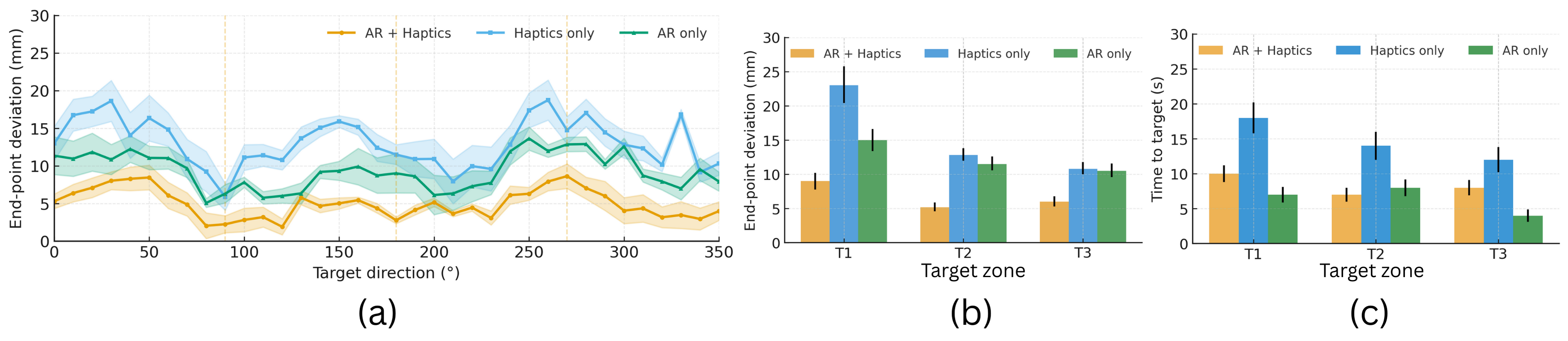}
    \caption{Experiment 2 results. (a) End-point deviation segmented by target direction (36 angular groups). The \textit{AR+Haptics} condition consistently yielded lower deviation across target directions. (b) End-point deviation grouped by radial target zone (T1, T2, T3). Across all conditions, targets in zone T1 (farthest) resulted in greater deviation than those in T2 and T3. (c) Time-to-target grouped by target zone. The \textit{Haptics-only} condition consistently required more time across all zones.}
    \label{fig:endpoint}
    \end{centering}
\end{figure*}

\subsubsection*{Procedure}
Participants completed a demographic questionnaire prior to the experiment. Each participant wore the vibrotactile wristband and the HoloLens 2, and held the same handheld tool used in Experiment~1. A knee phantom, \ADD{as shown in \cref{fig:phantom}a,} was fixed on a table, and participants were seated in a chair positioned in front of the phantom. A surgical light was projected onto the target to simulate surgical lighting conditions. Participants were instructed to maintain a consistent grip on the tool throughout the study. Our custom haptic system described in \cref{sec:4} was applied here. 

We defined 36 directions with 10$^{\circ}$ angular increments, beginning from the forward direction (0$^{\circ}$). The operating plane over the bone surface was subdivided into three radial zones originating from point $O$: T3 (nearest to the origin), T2 (intermediate), and T1 (farthest). Along each of the 36 directions, one target point was selected per zone, resulting in 108 unique target locations. At each target position, a 10~mm diameter semi-transparent sphere (50\% opacity) was rendered using Unity’s standard shader.

For each participant, 12 target points were randomly sampled from each of the three experimental conditions, yielding a total of 36 trials per participant. The order of the conditions was counterbalanced and randomly assigned across participants. HL2SS (\ADD{HoloLens 2 Sensor Streaming}) was used to enable the experiment coordinator to remotely initiate random target positions via a PC console \cite{dibene2022hololens}.

Participants completed 10 training trials to become familiar with both the haptic cue sequences and the AR visualizations. During each trial, participants positioned the tool tip at the origin of the bone plane \ADD{as shown in }\cref{fig:phantom}b. For conditions involving haptic guidance, the predefined haptic patterns (detailed in \cref{fig:4}) were automatically triggered based on the relative spatial relationship between the tool tip and the target. To enable this interaction, all spatial components were synchronized within the HoloLens world coordinate frame.

The handheld tool was continuously tracked by the HoloLens to provide real-time tool tip position data \cite{martin2023sttar}. Simultaneously, the 108 virtual target points were registered to the physical phantom using a markerless registration algorithm \cite{yang2025easyreg}. This algorithm used the depth camera to capture a point cloud of the real-world scene, \ADD{as shown in }\cref{fig:phantom}a, which was then matched to the predefined 3D mesh model. Once registration was complete, the forward direction (0$^{\circ}$) was defined in the HoloLens world coordinate system, and the angular difference between the current tool tip position and the target was continuously computed relative to this fixed reference. To account for potential spatial errors from SLAM drift, registration inaccuracies, or tool tracking noise, we designed the angular steps to be 10$^{\circ}$ (as detailed in \cref{sec:4.1}), providing tolerance to such errors.

The Euclidean distance between the tool tip and the target center was also calculated continuously to trigger corresponding haptic sequences (e.g., \textit{Pause} or \textit{Arrived} cues). In conditions involving AR visualization, the target sphere was directly rendered in the HMD, with or without concurrent haptic feedback. The participant’s task was to move the tool tip to the target center. Upon completion, participants pressed a Bluetooth-connected foot pedal to confirm the final tool position.

For each trial, we measured the \textbf{end-point deviation} (in mm), defined as the final distance between the tool tip and the target center minus 10~mm (the target radius). If the final distance was within 10~mm, the deviation was recorded as zero. We also recorded the \textbf{time-to-target}, defined as the duration from trial onset to foot pedal confirmation. Participants also completed a post-survey adopted from NASA-TLX for workload assessment, and the System Usability Scale (SUS) for usability measurements \cite{hart2006nasa, lewis2018system}. A semi-structured interview was also conducted at the end of experiment 2.

\subsubsection*{Results}
\textbf{End-Point Deviation.} A Shapiro-Wilk test indicated that the distribution of end-point deviation significantly deviated from normality ($p < .001$). As a result, we conducted a non-parametric Friedman test to compare performance across the three experimental conditions: \textit{AR+Haptics}, \textit{Haptics-only}, and \textit{AR-only}. The test revealed a statistically significant difference among the groups $p < .001$. 

Post-hoc pairwise comparisons using the Nemenyi test identified a significant difference between the \textit{AR+Haptics} and \textit{Haptics-only} conditions ($p < .01$), while the comparison between \textit{AR+Haptics} and \textit{AR-only} also reached statistical significance at a lower magnitude ($p < .05$). As illustrated in \cref{fig:endpoint}a, the \textit{AR+Haptics} condition yielded the lowest average end-point deviation (5.8~mm), followed by \textit{AR-only} (9.3~mm), and \textit{Haptics-only} (15.1~mm).

To further examine spatial performance, we grouped trials based on the 36 target directions. The general trend across directions confirmed that the \textit{AR+Haptics} condition consistently outperformed the \textit{Haptics-only} group, with reduced variance across angular positions. Additionally, as shown in ~\cref{fig:endpoint}b, when trials were grouped by target zone, a clear increase in mean deviation was observed in zone T1 (farthest from the tool origin) compared to T2 and T3, across all three conditions. While T2 and T3 yielded comparable performance, the T1 zone posed greater challenges, particularly for the \textit{Haptics-only} group. Notably, the mean deviation in the \textit{Haptics-only} group decreased by 11~mm when transitioning from T1 to T2, suggesting that distant targets are substantially more difficult to locate accurately using haptic feedback alone.

\textbf{Time-to-Target.} As the assumption of normality was met for time-to-target data (Shapiro-Wilk $p > .05$), we conducted a one-way repeated-measures ANOVA to compare completion times across the three experimental conditions. The analysis revealed a significant main effect of condition on time-to-target, $p < .001$. Post-hoc Bonferroni-adjusted comparisons showed that the \textit{Haptics-only} condition required significantly more time to reach the target ($M = 15.0$~s) compared to both the \textit{AR-only} and \textit{AR+Haptics} conditions ($p < .001$). 

Interestingly, the \textit{AR-only} condition yielded the shortest average completion time ($M = 6.3$~s), even outperforming the \textit{AR+Haptics} group ($M = 8.3$~s), despite the latter demonstrating the highest spatial precision. Minor fluctuations were observed across target zones: in the \textit{AR-only} group, T1 (the farthest zone) exhibited a slightly lower time-to-target than T2, and in the \textit{AR+Haptics} group, T3 (nearest zone) took marginally longer than T2.

Although we hypothesized that combining haptic guidance with AR would yield both the fastest and most accurate performance (H3), the results partially contradicted this assumption. While the \textit{AR+Haptics} group achieved the highest spatial precision, it did not produce the lowest time-to-target. In fact, participants in the \textit{AR-only} condition completed the task the fastest. One possible explanation is that the addition of haptic feedback increased cognitive load or divided attention, thereby slowing down motor response even as it improved end-point accuracy. Another plausible factor is that participants may have relied predominantly on the highly visible AR cues, with haptic feedback serving only as secondary reinforcement, thus increasing deliberation time before confirming target acquisition. These findings suggest that while multimodal cues can improve handheld tool movement and maneuver precision, they may not always enhance efficiency.

\textbf{Subjective Workload.}
We evaluated perceived workload using NASA-TLX composite scores (0--100), averaging the six standard subscales (mental demand, physical demand, temporal demand, performance, effort, and frustration). Lower values represent reduced load.

\begin{figure}[t]
  \centering
  \includegraphics[width=\linewidth]{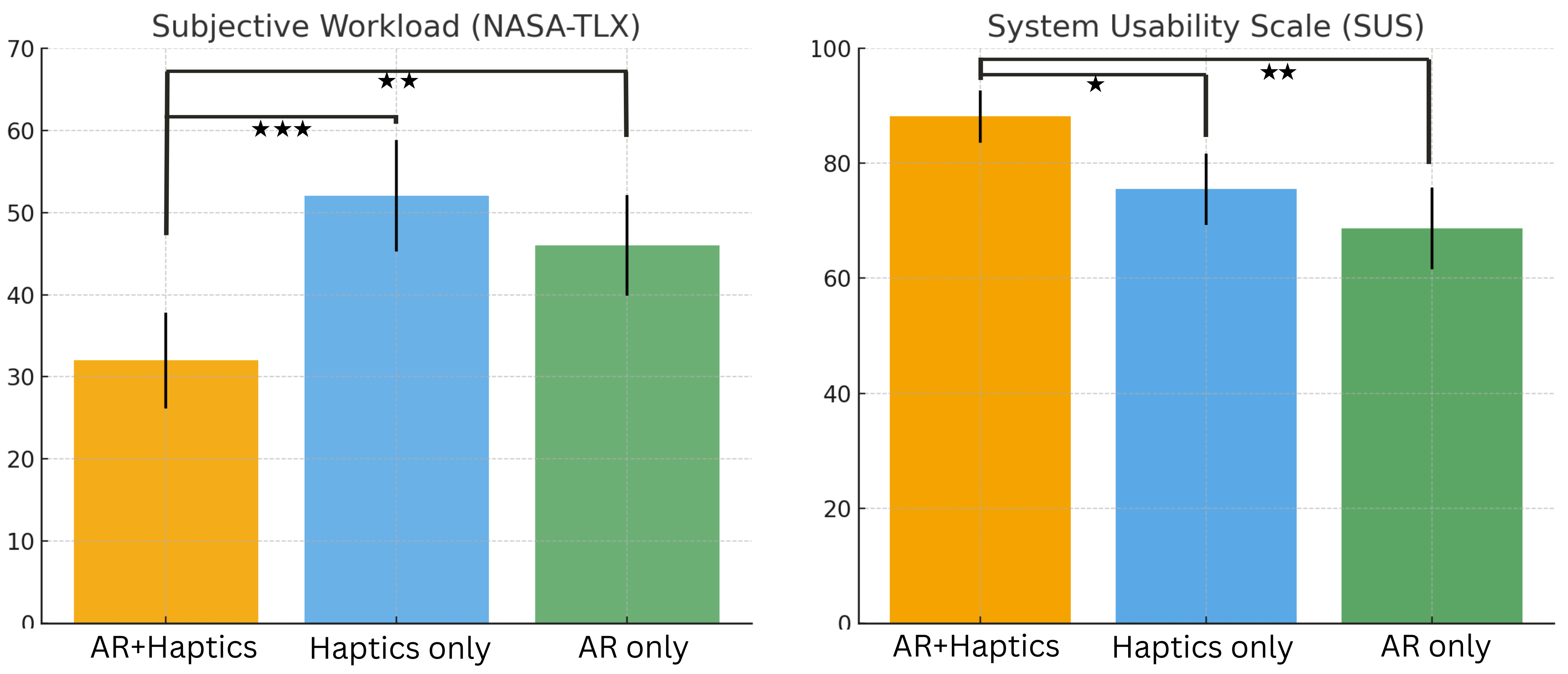}
  \caption{Subjective usability and workload results. Left: average NASA-TLX workload scores across conditions (lower is better). Right: System Usability Scale (SUS) across conditions (higher is better). Error bars indicate standard deviations. Asterisks denote significance levels: * $p < .05$, ** $p < .01$, *** $p < .001$.}
  \label{fig:workload}
\end{figure}

As shown in \cref{fig:workload} left, a Friedman test revealed a significant main effect of condition on workload ($\chi^2$(2) = 16.18, $p < .001$). Post-hoc Dunn’s tests with Bonferroni correction confirmed that \textit{AR+Haptics} ($M = 32.0$, $SD = 5.8$) elicited significantly lower workload than both \textit{AR-only} ($M = 46.0$, $SD = 6.1$, $p < .01$) and \textit{Haptics-only} ($M = 52.1$, $SD = 6.8$, $p < .001$). Interestingly, the difference between AR-only and Haptics-only remained non-significant ($p = .082$).

Participants consistently highlighted the complementary nature of visual and tactile feedback in the \textit{AR+Haptics} condition. Many described AR as providing coarse spatial orientation, with haptics cues feeling as "more confident." \textit{P4} mentioned, "The AR circle helps you get close, but it’s the buzz that tells you you’re exactly there. I really relied on the vibration to finish the job." \textit{P1} also said, "It felt like a GPS. You see where you're going, and the haptics guides you there and say you’ve arrived." Around 80\% of users also described difficulty in judging depth and alignment in the \textit{AR-only} condition: ``The dot looks helpful at first, but getting the tool aligned in 3D space is tiring under such a bright light [surgical light]. Sometimes I just wasn’t sure if I was actually on target.''- \textit{P6}

By contrast, \textit{Haptics-only} was frequently cited as cognitively demanding. Without visual guidance, participants needed to interpret cues and make fine motor adjustments based solely on vibration feedback, especially when task was just launched. ``When relying on haptics, it felt like trying to solve a puzzle.''-\textit{P3}. Another participant (\textit{P7}) also mentioned "I found it harder to keep track of where I was, and sometimes I am confused which direction I should go." This subjective burden aligned with elevated TLX scores and longer completion times observed in Haptics-only trials. However, several participants reported minimal problem relying on haptics cues, complementing the intuitive sequence design and easy-to-follow cues. Due to the planar nature of the task, vertical directional cues (e.g., ``move up'' or ``down'') were rarely triggered. As P2 noted, ``It was mostly a flat surface. I barely needed to go up or down unless I drifted too far.''-\textit{P2}

These findings support the hypothesis that \textit{AR+Haptics} significantly reduces cognitive workload by combining intuitive visual orientation with confirmatory tactile feedback. While \textit{AR-only} improves targeting compared to haptics alone, it struggles under depth ambiguity and limitations in occlusion under surgical light conditions, despite having a relatively small target visualization. Compared with \textit{AR-only}, \textit{Haptics-only} results in a higher workload, while this difference is not statistically significant. 

\textbf{System Usability}
A Friedman test revealed a statistically significant effect of condition on System Usability Scale (SUS) scores ($p < .01$). \ADD{Post-hoc pairwise comparisons using Dunn’s test with Bonferroni correction indicated that} the \textit{AR+Haptics} condition ($M = 88.1$, $SD = 4.5$) was rated significantly more usable than both the \textit{Haptics-only} condition ($M = 75.5$, $SD = 6.2$, $p < .05$) and the \textit{AR-only} condition ($M = 68.7$, $SD = 7.1$, $p < .01$). While Haptics-only marginally outperformed AR-only, this difference did not reach statistical significance, suggesting that neither unimodal interface alone provided the highest usability for surgical-like environments.

Qualitative feedback corroborated these quantitative results. Participants described the \textit{AR+Haptics} condition as "seamless," "easy to learn," and "the most natural to use." One noted that "it felt like having a mutual backup system, and vice versa." The combination of visual confirmation and tactile guidance was repeatedly praised as "intuitive," with multiple participants indicating it provided "confidence" and minimized ambiguity during alignment tasks. Another participant remarked, "it feels like the system was working with me, not against me."

In contrast, the \textit{AR-only} condition was criticized for its limitations under suboptimal lighting conditions. Several participants noted that "the tooltip became hard to see when the room was too bright with the semi-transparent circle [target]." Others expressed frustration over relying solely on visual cues in a depth-constrained, see-through display environment. One participant stated, "without any tactile confirmation, I was constantly second-guessing whether I had reached the right target."

The Haptics-only condition was perceived as moderately usable but cognitively demanding, thus partially contradicting H4. Participants described the vibration patterns as "clear but mentally tiring to interpret, especially all the time." While some appreciated the tactile feedback, others mentioned that "having to subconsciously remember what each vibration meant slowed me down." Despite these limitations, Haptics-only still outperformed AR-only in mean SUS scores, which may reflect the reliability and consistency of tactile cues in contrast to visually ambiguous overlays.

Overall, the combined modality was not only rated highest in usability but was also consistently described as the most trustworthy and user-friendly configuration. These results support the integration of multimodal feedback as a practical approach to enhancing usability in mixed-reality surgical environments, and the high usability of our haptic system provides important design guidance. 

\section{Discussion}
\subsection*{Multimodal Guidance Improves Accuracy but Trades Off Speed}
Our findings suggest that combining wrist-based haptics with AR visual overlays can significantly improve spatial precision in handheld tool tasks. Participants in the \textit{AR+Haptics} condition consistently achieved the lowest end-point deviation ($M = 5.8~mm$), outperforming both \textit{AR-only} and \textit{Haptics-only} conditions. This confirms the benefit of multimodal integration for spatial guidance, aligning with prior work on sensory redundancy in navigation systems \cite{arbelaez2019haptic, zhang2023towards}. \ADD{Our system also satisfies the level of accuracy needed for clinical use \cite{jeung2025augmented}.}

However, the improved accuracy came at the cost of longer time-to-target compared to \textit{AR-only}. While this partially contradicts our hypothesis (H3), it reflects a critical trade-off: although haptic cues enhanced spatial certainty, they also introduced a secondary channel of information that users had to attend to and interpret. As one participant noted, "I was more confident with the vibration, but it slowed me down a bit because I waited to feel it." This tension underscores the importance of designing multimodal systems that balance confidence with efficiency, especially given the surgical context where both precision and time matter.

\subsection*{Visual Dominance vs. Haptic Reassurance}
Participant feedback revealed clear differences in how each modality shaped interaction. AR visuals were often described as "quick but unreliable," particularly under challenging lighting conditions or depth ambiguity. Several users stated that "the dot looked helpful, but I wasn\'t sure if I was really at the target." In contrast, the haptic feedback was perceived as "trustworthy" and "reassuring," especially in later stages of tool placement.

This supports a model where AR provides coarse targeting, and haptics refines final placement---a division of labor that can inform future guidance systems. Rather than competing, the two modalities can be optimized to serve complementary roles in phased interactions: vision for global orientation, and touch for local confirmation.

\subsection*{Cognitive Load and Perceived Workload}
Despite AR-only yielding faster task times, its workload scores were significantly higher than \textit{AR+Haptics}. Participants reported greater mental effort in aligning the tool using visuals alone, with one noting, "I had to keep guessing if I was too deep or too far." Conversely, \textit{Haptics-only} caused the highest workload and lowest performance, likely due to the absence of external spatial anchors. Users often described the experience as "mentally tiring" or "like solving a puzzle blindly."

These patterns align with cognitive offloading theories: tactile cues offload mental processing by externalizing feedback through bodily sensation \cite{pezent2019tasbi, hessinger2020wearable}. Our results suggest that carefully designed haptic cues can mitigate visual overload and enhance user confidence---key for high-stakes, visually demanding tasks such as surgery.

\subsection*{Design Implications for Surgical Interfaces}
Our findings suggest several takeaways for designers of AR-guided surgical systems: 1) Designers should prioritize when and how to introduce tactile cues depending on the task phase (e.g., alignment vs. confirmation); 2) In settings with occlusion, lighting, or depth ambiguity, tactile reinforcement provides crucial redundancy; 3) Our custom-designed vibration sequence, incorporating phantom interpolation and urgency-based pulse encoding, contributed to high cue recognition rates, even under visual interference. These lessons may apply not only to surgical AR systems but also to any application where hand-held precision tasks occur under sensory or cognitive load.

\subsection*{Limitations and Future Work}
This study was conducted in a controlled laboratory setting using a simulated bone phantom and optical see-through AR. Future work should evaluate the system in real clinical environments, where distractions, occlusion, and stressors are more complex. The haptic system was limited to wrist-level cues; expanding to finger or tool-integrated feedback may improve spatial specificity. Additionally, future systems could adapt cue intensity or frequency dynamically based on user performance or error trends, implementing intelligent feedback control. \ADD{Future work should include more expert surgeons in all evaluation phases to validate our findings.}

\ADD{While our results demonstrate statistically significant gains in accuracy and subjective metrics, certain results are subject to overestimation due to pooled data, and long-term training effects remain open questions.} As one participant noted, "The haptics grew on me—it took some getting used to, but then it clicked." Investigating learning curves, fatigue, and trust over time will be critical for clinical adoption.

\section{Conclusion}
We presented a multimodal handheld guidance system that combines wrist-based vibrotactile cues with optical see-through augmented reality (AR) to support spatial precision tasks such as surgical tool alignment. Our work contributes a full design and validation pipeline—from formative user elicitation, to cue recognition evaluation, to closed-loop guidance experiments. A formative study with experienced surgeons and residents revealed strong user preference for a fixed anatomical reference frame over adaptive tool-referenced mappings, informing our directional encoding scheme. Participants also identified a small, meaningful set of tool maneuvers that guided our pattern design.

In Experiment~1, participants demonstrated high recognition accuracy for all five cue types, with significantly better performance for full-actuator cues (e.g., \textit{Pause} and \textit{Arrived}) over more spatially localized directional patterns. These findings validate the design of our vibration sequences under visual AR overlay, and suggest that multi-motor activation can improve cue salience, even in attentionally demanding environments.

In Experiment~2, we evaluated tool movement performance across \textit{Haptics-only}, \textit{AR-only}, and \textit{AR+Haptics} conditions. Results showed that combining haptic and visual feedback significantly improved spatial precision and reduced cognitive workload compared to either modality alone. While \textit{AR-only} yielded the fastest task times, it was also rated lowest in usability metrics, revealing an interesting trade-off. Participants reported that the haptic cues offered reassuring confirmation, especially under visual ambiguity and lighting challenges.

These results show that well-designed wrist-based haptics can effectively complement AR by providing non-visual directional guidance and reducing reliance on continuous visual attention. Our work highlights the value of multimodal feedback in surgery-like tasks and offers practical design insights for wearable haptic interfaces in AR. Future directions include evaluating adaptive cue systems, exploring finger-based actuation, and conducting in-situ testing in operating room environments to assess clinical transferability.

\section{Declarations}
This study was approved by the Stanford research ethics committee, and consent from participants was obtained.

\bibliographystyle{abbrv-doi-hyperref-narrow}

\bibliography{template}

@article{birkmeyer2003surgeon,
  title={Surgeon volume and operative mortality in the United States},
  author={Birkmeyer, John D and Stukel, Therese A and Siewers, Andrea E and Goodney, Philip P and Wennberg, David E and Lucas, F Lee},
  journal={New England Journal of Medicine},
  volume={349},
  number={22},
  pages={2117--2127},
  year={2003},
  publisher={Mass Medical Soc}
}

@article{halm2002volume,
  title={Is volume related to outcome in health care? A systematic review and methodologic critique of the literature},
  author={Halm, Ethan A and Lee, Clara and Chassin, Mark R},
  journal={Annals of internal medicine},
  volume={137},
  number={6},
  pages={511--520},
  year={2002},
  publisher={American College of Physicians}
}

@article{kelz2021national,
  title={A national comparison of operative outcomes of new and experienced surgeons},
  author={Kelz, Rachel R and Sellers, Morgan M and Niknam, Bijan A and Sharpe, James E and Rosenbaum, Paul R and Hill, Alexander S and Zhou, Hong and Hochman, Lauren L and Bilimoria, Karl Y and Itani, Kamal and others},
  journal={Annals of surgery},
  volume={273},
  number={2},
  pages={280--288},
  year={2021},
  publisher={LWW}
}

@article{kraus2025fellowship,
  title={Fellowship-trained surgeons experience a learning curve performing revision total joint arthroplasty},
  author={Kraus, Kent R and Harris, Alexander C and Ziemba-Davis, Mary and Buller, Leonard T and Meneghini, R Michael},
  journal={The Journal of Arthroplasty},
  volume={40},
  number={1},
  pages={28--33},
  year={2025},
  publisher={Elsevier}
}

@article{nema2022surgical,
  title={Surgical instrument detection and tracking technologies: Automating dataset labeling for surgical skill assessment},
  author={Nema, Shubhangi and Vachhani, Leena},
  journal={Frontiers in Robotics and AI},
  volume={9},
  pages={1030846},
  year={2022},
  publisher={Frontiers Media SA}
}

@inproceedings{tsui2023optical,
  title={An optical tracking approach to computer-assisted surgical navigation via stereoscopic vision},
  author={Tsui, Darin and Melentyev, Capalina and Rajan, Ananya and Kumar, Rohan and Talke, Frank E},
  booktitle={Information Storage and Processing Systems},
  volume={87219},
  pages={V001T05A003},
  year={2023},
  organization={American Society of Mechanical Engineers}
}

@article{ghazanfar2015effect,
  title={The effect of divided attention on novices and experts in laparoscopic task performance},
  author={Ghazanfar, Mudassar Ali and Cook, Malcolm and Tang, Benjie and Tait, Iain and Alijani, Afshin},
  journal={Surgical endoscopy},
  volume={29},
  number={3},
  pages={614--619},
  year={2015},
  publisher={Springer}
}

@article{van2024augmented,
  title={Augmented reality guidance improves accuracy of orthopedic drilling procedures},
  author={Van Gestel, Frederick and Van Aerschot, Fiene and Frantz, Taylor and Verhellen, Anouk and Barb{\'e}, Kurt and Jansen, Bart and Vandemeulebroucke, Jef and Duerinck, Johnny and Scheerlinck, Thierry},
  journal={Scientific Reports},
  volume={14},
  number={1},
  pages={25269},
  year={2024},
  publisher={Nature Publishing Group UK London}
}

@article{park2020augmented,
  title={Augmented reality improves procedural efficiency and reduces radiation dose for CT-guided lesion targeting: a phantom study using HoloLens 2},
  author={Park, Brian J and Hunt, Stephen J and Nadolski, Gregory J and Gade, Terence P},
  journal={Scientific reports},
  volume={10},
  number={1},
  pages={18620},
  year={2020},
  publisher={Nature Publishing Group UK London}
}

@article{yang2025easyreg,
  title={EasyREG: Easy Depth-Based Markerless Registration and Tracking using Augmented Reality Device for Surgical Guidance},
  author={Yang, Yue and Leuze, Christoph and Hargreaves, Brian and Daniel, Bruce and Baik, Fred},
  journal={arXiv preprint arXiv:2504.09498},
  year={2025}
}

@article{van2022high,
  title={High-Accuracy Augmented Reality Guidance for Intracranial Drain Placement Using a Standalone Head-Worn Navigation System: First-in-Human Results},
  author={Van Gestel, Frederick and Frantz, Taylor and Buyck, F{\'e}lix and Gallagher, Anthony G and Geens, Wietse and Neuville, Quentin and Bruneau, Michael and Jansen, Bart and Scheerlinck, Thierry and Vandemeulebroucke, Jef and others},
  journal={Neurosurgery},
  pages={10--1227},
  year={2022},
  publisher={LWW}
}

@article{condino2019perceptual,
  title={Perceptual limits of optical see-through visors for augmented reality guidance of manual tasks},
  author={Condino, Sara and Carbone, Marina and Piazza, Roberta and Ferrari, Mauro and Ferrari, Vincenzo},
  journal={IEEE Transactions on Biomedical Engineering},
  volume={67},
  number={2},
  pages={411--419},
  year={2019},
  publisher={IEEE}
}

@phdthesis{clinical,
  title={Clinical Application to Improve the “Depth Perception Problem” by Combining Augmented Reality and a 3D Printing Model},
  author={pianmeishanli},
  school={Osaka Medical and Pharmaceutical University}
}

@article{doughty2022augmenting,
  title={Augmenting performance: A systematic review of optical see-through head-mounted displays in surgery},
  author={Doughty, Mitchell and Ghugre, Nilesh R and Wright, Graham A},
  journal={Journal of Imaging},
  volume={8},
  number={7},
  pages={203},
  year={2022},
  publisher={MDPI}
}

@article{muzzammil2024review,
  title={A review on tissue-needle interaction and path planning models for bevel tip type flexible needle minimal intervention},
  author={Muzzammil, Hafiz Muhammad and Zhang, Yong-De and Ejaz, Hassan and Yuan, Qihang and Muddassir, Muhammad},
  journal={Math Biosci Eng},
  volume={21},
  number={1},
  pages={523--61},
  year={2024}
}

@article{aggravi2021haptic,
  title={Haptic teleoperation of flexible needles combining 3d ultrasound guidance and needle tip force feedback},
  author={Aggravi, Marco and Estima, Daniel AL and Krupa, Alexandre and Misra, Sarthak and Pacchierotti, Claudio},
  journal={IEEE Robotics and automation letters},
  volume={6},
  number={3},
  pages={4859--4866},
  year={2021},
  publisher={IEEE}
}

@article{rossa2016hand,
  title={A hand-held assistant for semiautomated percutaneous needle steering},
  author={Rossa, Carlos and Usmani, Nawaid and Sloboda, Ronald and Tavakoli, Mahdi},
  journal={IEEE Transactions on Biomedical Engineering},
  volume={64},
  number={3},
  pages={637--648},
  year={2016},
  publisher={IEEE}
}

@article{yamamoto2012augmented,
  title={Augmented reality and haptic interfaces for robot-assisted surgery},
  author={Yamamoto, Tomonori and Abolhassani, Niki and Jung, Sung and Okamura, Allison M and Judkins, Timothy N},
  journal={The International Journal of Medical Robotics and Computer Assisted Surgery},
  volume={8},
  number={1},
  pages={45--56},
  year={2012},
  publisher={Wiley Online Library}
}

@article{raitor2023design,
  title={Design and Evaluation of Haptic Guidance in Ultrasound-Based Needle-Insertion Procedures},
  author={Raitor, Michael and Nunez, Cara M and Stolka, Philipp J and Okamura, Allison M and Culbertson, Heather},
  journal={IEEE Transactions on Biomedical Engineering},
  volume={71},
  number={1},
  pages={26--35},
  year={2023},
  publisher={IEEE}
}

@article{li2022review,
  title={A review on the techniques used in prostate brachytherapy},
  author={Li, Yanlei and Yang, Chenguang and Bahl, Amit and Persad, Raj and Melhuish, Chris},
  journal={Cognitive Computation and Systems},
  volume={4},
  number={4},
  pages={317--328},
  year={2022},
  publisher={Wiley Online Library}
}

@article{bergholz2023benefits,
  title={The benefits of haptic feedback in robot assisted surgery and their moderators: a meta-analysis},
  author={Bergholz, Max and Ferle, Manuel and Weber, Bernhard M},
  journal={Scientific Reports},
  volume={13},
  number={1},
  pages={19215},
  year={2023},
  publisher={Nature Publishing Group UK London}
}

@inproceedings{hong2016evaluating,
  title={Evaluating angular accuracy of wrist-based haptic directional guidance for hand movement.},
  author={Hong, Jonggi and Stearns, Lee and Froehlich, Jon and Ross, David and Findlater, Leah},
  booktitle={Graphics Interface},
  pages={195--200},
  year={2016}
}

@article{elsayed2020vibromap,
  title={Vibromap: Understanding the spacing of vibrotactile actuators across the body},
  author={Elsayed, Hesham and Weigel, Martin and M{\"u}ller, Florian and Schmitz, Martin and Marky, Karola and G{\"u}nther, Sebastian and Riemann, Jan and M{\"u}hlh{\"a}user, Max},
  journal={Proceedings of the ACM on Interactive, Mobile, Wearable and Ubiquitous Technologies},
  volume={4},
  number={4},
  pages={1--16},
  year={2020},
  publisher={ACM New York, NY, USA}
}

@article{wang2018empirical,
  title={An empirical evaluation on vibrotactile feedback for wristband system},
  author={Wang, Feng and Zhang, Wanna and Luo, Wei},
  journal={Mobile Information Systems},
  volume={2018},
  number={1},
  pages={4878014},
  year={2018},
  publisher={Wiley Online Library}
}

@misc{salazar2018path,
  title={Path-Following Guidance Using Phantom Sensation Based Vibrotactile Cues Around the Wrist. IEEE Robotics and Automation Letters 3, 3 (2018), 2485--2492},
  author={Salazar, Jose and Okabe, Keisuke and Hirata, Yasuhisa},
  year={2018}
}

@article{rossa2016multiactuator,
  title={Multiactuator haptic feedback on the wrist for needle steering guidance in brachytherapy},
  author={Rossa, Carlos and Fong, Jason and Usmani, Nawaid and Sloboda, Ronald and Tavakoli, Mahdi},
  journal={IEEE Robotics and Automation Letters},
  volume={1},
  number={2},
  pages={852--859},
  year={2016},
  publisher={IEEE}
}

@inproceedings{hessinger2020wearable,
  title={Wearable Vibrotactile Interface Using Phantom Tactile Sensation for Human-Robot Interaction},
  author={Hessinger, Markus and Beckerle, Philipp and Kupnik, Mario},
  booktitle={Haptics: Science, Technology, Applications: 12th International Conference, EuroHaptics 2020, Leiden, The Netherlands, September 6--9, 2020, Proceedings},
  volume={12272},
  pages={380},
  year={2020},
  organization={Springer Nature}
}

@article{zhang2023towards,
  title={Towards reducing visual workload in surgical navigation: proof-of-concept of an augmented reality haptic guidance system},
  author={Zhang, Gesiren and Bartels, Jan and Martin-Gomez, Alejandro and Armand, Mehran},
  journal={Computer Methods in Biomechanics and Biomedical Engineering: Imaging \& Visualization},
  volume={11},
  number={4},
  pages={1073--1080},
  year={2023},
  publisher={Taylor \& Francis}
}

@article{arbelaez2019haptic,
  title={Haptic augmented reality (HapticAR) for assembly guidance},
  author={Arbel{\'a}ez, JC and Vigan{\`o}, Roberto and Osorio-G{\'o}mez, Gilberto},
  journal={International Journal on Interactive Design and Manufacturing (IJIDeM)},
  volume={13},
  number={2},
  pages={673--687},
  year={2019},
  publisher={Springer}
}

@inproceedings{pezent2019tasbi,
  title={Tasbi: Multisensory squeeze and vibrotactile wrist haptics for augmented and virtual reality},
  author={Pezent, Evan and Israr, Ali and Samad, Majed and Robinson, Shea and Agarwal, Priyanshu and Benko, Hrvoje and Colonnese, Nick},
  booktitle={2019 IEEE World Haptics Conference (WHC)},
  pages={1--6},
  year={2019},
  organization={IEEE}
}

@article{lee2021wearable,
  title={Wearable haptic device for stiffness rendering of virtual objects in augmented reality},
  author={Lee, Yongseok and Lee, Somang and Lee, Dongjun},
  journal={Applied Sciences},
  volume={11},
  number={15},
  pages={6932},
  year={2021},
  publisher={MDPI}
}

@article{bhatia2024augmenting,
  title={Augmenting the feel of real objects: An analysis of haptic augmented reality},
  author={Bhatia, Arpit and Hornb{\ae}k, Kasper and Seifi, Hasti},
  journal={International Journal of Human-Computer Studies},
  volume={185},
  pages={103244},
  year={2024},
  publisher={Elsevier}
}

@article{Wang_Wang_Ren_2024, title={Visual and haptic guidance for enhancing target search performance in dual-task settings}, volume={14}, number={11}, journal={Applied Sciences}, author={Wang, Gang and Wang, Hung-Hsiang and Ren, Gang}, year={2024}, month={May}, pages={4650}}

@inproceedings{scheggi2014vibrotactile,
  title={A vibrotactile bracelet to improve the navigation of older adults in large and crowded environments},
  author={Scheggi, Stefano and Aggravi, Marco and Prattichizzo, Domenico and others},
  booktitle={Proc. 20th IMEKO TC4 Int. Symp. and 18th Int. Workshop on ADC Modelling and Testing Research on Electric and Electronic Measurement for the Economic Upturn},
  pages={798--801},
  year={2014}
}

@misc{Industries, title={Vibrating Mini Motor disc}, url={https://www.adafruit.com/product/1201}, journal={adafruit industries blog RSS}, author={Industries, Adafruit}}

@article{martinez2022psychophysical,
  title={Psychophysical Studies of Interleaving Narrowband Tactile Stimuli to Achieve Broadband Perceptual Effects},
  author={Martinez, Juan S and Tan, Hong Z and Cholewiak, Roger W},
  journal={Frontiers in Virtual Reality},
  volume={3},
  pages={894575},
  year={2022},
  publisher={Frontiers Media SA}
}

@article{martin2023sttar,
  title={Sttar: surgical tool tracking using off-the-shelf augmented reality head-mounted displays},
  author={Martin-Gomez, Alejandro and Li, Haowei and Song, Tianyu and Yang, Sheng and Wang, Guangzhi and Ding, Hui and Navab, Nassir and Zhao, Zhe and Armand, Mehran},
  journal={IEEE Transactions on Visualization and Computer Graphics},
  volume={30},
  number={7},
  pages={3578--3593},
  year={2023},
  publisher={IEEE}
}

@inproceedings{kato2010basic,
  title={Basic properties of phantom sensation for practical haptic applications},
  author={Kato, Hiroshi and Hashimoto, Yuki and Kajimoto, Hiroyuki},
  booktitle={International Conference on Human Haptic Sensing and Touch Enabled Computer Applications},
  pages={271--278},
  year={2010},
  organization={Springer}
}

@article{alles1970information,
  title={Information transmission by phantom sensation},
  author={Alles, David S},
  journal={IEEE Trans. Man-Machine Syst.},
  volume={11},
  number={1},
  pages={28--35},
  year={1970}
}

@article{remache2024phantom,
  title={Phantom sensation: Threshold and quality indicators of a tactile illusion of motion},
  author={Remache-Vinueza, Byron and Trujillo-Le{\'o}n, Andr{\'e}s and Vidal-Verd{\'u}, Fernando},
  journal={Displays},
  volume={83},
  pages={102676},
  year={2024},
  publisher={Elsevier}
}

@inproceedings{rahal2009investigating,
  title={Investigating the influence of temporal intensity changes on apparent movement phenomenon},
  author={Rahal, Lara and Cha, Jongeun and El Saddik, Abdulmotaleb and Kammerl, Julius and Steinbach, Eckehard},
  booktitle={2009 IEEE International Conference on Virtual Environments, Human-Computer Interfaces and Measurements Systems},
  pages={310--313},
  year={2009},
  organization={IEEE}
}

@article{schafer2025vibrotactile,
  title={Vibrotactile Phantom Sensations in Haptic Wrist Rotation Guidance},
  author={Sch{\"a}fer, Niklas and Seiler, Julian and Latsch, Bastian and Kupnik, Mario and Beckerle, Philipp},
  journal={IEEE Transactions on Haptics},
  year={2025},
  publisher={IEEE}
}

@article{kirchner2023phantom,
  title={Phantom illusion based vibrotactile rendering of affective touch patterns},
  author={Kirchner, Robert and Rosenkranz, Robert and Sousa, Brais Gonzalez and Li, Shu-Chen and Altinsoy, M Ercan},
  journal={IEEE Transactions on Haptics},
  volume={17},
  number={2},
  pages={202--215},
  year={2023},
  publisher={IEEE}
}

@article{dibene2022hololens,
  title={HoloLens 2 Sensor Streaming},
  author={Dibene, Juan C and Dunn, Enrique},
  journal={arXiv preprint arXiv:2211.02648},
  year={2022}
}

@inproceedings{hart2006nasa,
  title={NASA-task load index (NASA-TLX); 20 years later},
  author={Hart, Sandra G},
  booktitle={Proceedings of the human factors and ergonomics society annual meeting},
  volume={50},
  number={9},
  pages={904--908},
  year={2006},
  organization={Sage publications Sage CA: Los Angeles, CA}
}

@article{lewis2018system,
  title={The system usability scale: past, present, and future},
  author={Lewis, James R},
  journal={International Journal of Human--Computer Interaction},
  volume={34},
  number={7},
  pages={577--590},
  year={2018},
  publisher={Taylor \& Francis}
}

@article{jeung2025augmented,
  title={Augmented Reality With Dynamic Anatomy Modelling for Knee Arthroscopy},
  author={Jeung, Deokgi and Lee, Hyun-Joo and Kim, Hee-June and Choi, Hyunseok and Hong, Jaesung},
  journal={Healthcare Technology Letters},
  volume={12},
  number={1},
  pages={e70034},
  year={2025},
  publisher={Wiley Online Library}
}

@article{erp2005waypoint,
  title={Waypoint navigation with a vibrotactile waist belt},
  author={Erp, Jan BF Van and Veen, Hendrik AHC Van and Jansen, Chris and Dobbins, Trevor},
  journal={ACM Transactions on Applied Perception (TAP)},
  volume={2},
  number={2},
  pages={106--117},
  year={2005},
  publisher={ACM New York, NY, USA}
}
\clearpage
\onecolumn
\listofchanges
\clearpage
\end{document}